\documentclass{iopart}
\pdfoutput=1
\usepackage{iopams}  
\usepackage{graphicx}

\begin{document}

\title[High-frequency corrections to the detector response]
{High-frequency corrections to the detector response and their effect 
on searches for gravitational waves}

\author{M Rakhmanov${}^{1}$, J D Romano${}^{1,2}$ and J T Whelan${}^3$}

\address{${}^1$ University of Texas at Brownsville, Brownsville, Texas, 
78520, USA}
\address{${}^2$ Cardiff University, Cardiff CF24 3AA, Wales, UK}
\address{${}^3$ Max-Planck-Institut f\"{u}r Gravitationsphysik
  (Albert-Einstein-Institut), D-14476 Potsdam, Germany}

\eads{
\mailto{malik@phys.utb.edu}, 
\mailto{joe@phys.utb.edu}, 
\mailto{john.whelan@aei.mpg.de}
}

\begin{abstract}
Searches for gravitational waves with km-scale laser interferometers
often involve the long-wavelength approximation to describe the
detector response. The prevailing assumption is that the corrections
to the detector response due to its finite size are small
and the errors due to the long-wavelength approximation are negligible. 
Recently, however, Baskaran and Grishchuk 
(2004 {\em Class.\ Quantum Grav.}\ {\bf 21} 4041) found 
that in a simple Michelson interferometer such errors can be as large
as 10 percent. For more accurate analysis, these authors proposed to
use a linear-frequency correction to the long-wavelength approximation.
In this paper we revisit these calculations. We show that the 
linear-frequency correction is inadequate for certain locations in the 
sky and therefore accurate analysis requires taking into account the 
exact formula, commonly derived from the photon round-trip propagation
time. Also, we extend the calculations to include the effect of
Fabry-Perot resonators in the interferometer arms. Here we show that 
a simple approximation which combines the long-wavelength Michelson
response with the single-pole approximation to the Fabry-Perot
transfer function produces rather accurate results. In particular, the 
difference between the exact and the approximate formulae is at most 
2-3 percent for those locations in the sky where the detector response 
is greater than half of its maximum value. We analyse the impact of
such errors on detection sensitivity and parameter estimation in 
searches for periodic gravitational waves emitted by a known pulsar,
and in searches for an isotropic stochastic gravitational-wave
background. At frequencies up to 1~kHz, the effect of such
errors is at most 1-2 percent. For higher frequencies, or if more 
accuracy is required, one should use the exact formula 
for the detector response.
\end{abstract}


\section{Introduction}

Searches for gravitational waves are currently conducted with km-scale 
laser interferometers such as LIGO \cite{Barish:1999} and VIRGO 
\cite{Bradaschia:1990}. These detectors utilize a Michelson
configuration which is further enhanced by the addition of Fabry-Perot
cavities in the interferometer arms. Development of efficient data
analysis algorithms requires accurate knowledge of the response of
these detectors to gravitational waves. There are two somewhat 
different points of view 
on how to calculate the detector response. In one approach,
it is assumed that the size of the detector is much less than the
wavelength of the incoming gravitational wave, and therefore can be 
neglected in the calculations. This approach is often called the 
\emph{long-wavelength} approximation \cite{Misner:1973, Thorne:1987}. 
The advantage of this approximation is that it allows one to interpret 
the effect of gravitational waves entirely in terms of the motion of
test masses, which is appealing to our physical intuition. 
In another approach, one takes into account the finite size of the 
detector by considering variations in the gravitational wave within 
the duration of one photon round trip between the test masses 
%\cite{Gursel:1984, Meers:1989b, Christensen:PhD, Christensen:1992, 
%Sigg:1997, Mizuno:1997}. 
\cite{Gursel:1984}-\cite{Mizuno:1997}. 
The detector response obtained in this way is more accurate but 
no longer allows the simple interpretation 
in terms of test mass motion \cite{Rakhmanov:2005}. Such calculations 
are commonly used to derive the response of space-borne  
gravitational-wave antennae 
%\cite{Estabrook:1975, Estabrook:1985, 
%Schilling:1997, Larson:2000, Cornish:2003}. 
\cite{Estabrook:1975}-\cite{Cornish:2003}. 
For ground-based
detectors, one usually adopts the long-wavelength approximation,
assuming that it is sufficiently accurate.

It was pointed out by Baskaran and Grishchuk \cite{Baskaran:2004} 
that even for ground-based detectors, the long-wavelength approximation 
can lead to noticeable errors in the estimation of parameters of a  
gravitational wave. In particular, they found that the error in 
searches for periodic gravitational waves can be as large as 10\%,
thus raising a concern about the validity of the long-wavelength
approximation in recent searches for gravitational waves. Some 
key points of their analysis however required clarification.  
The authors assumed that for ground-based detectors it suffices to use
the first-order correction to the long-wavelength approximation and 
thus introduced a linear-frequency detector response, whereas the
exact, non-linear formula was readily available. Also, the
calculations did not take into account the presence of Fabry-Perot 
cavities in the interferometer arms, which play a crucial role in 
the formation of the signal. It is therefore worthwhile to reconsider 
this analysis.

In this paper we re-evaluate the errors due to the long-wavelength 
approximation and assess their impact on current searches for 
gravitational waves with km-scale laser interferometers. We show that 
the linear-frequency approximation is inadequate for some locations 
in the sky, and therefore one must use the exact formula for the
detector response to estimate systematic errors from the
long-wavelength approximation. To make the analysis applicable for LIGO
and VIRGO detectors, we include the transfer function of Fabry-Perot
cavities in the interferometer arms. Using the exact expression for
the detector response, we estimate the errors resulting from the 
long-wavelength approximation in searches for periodic gravitational 
waves and in searches for an isotropic stochastic gravitational-wave 
background.

\section{Michelson interferometer response (long-wavelength 
approximation)}
\label{s:antenna-LW}

In the transverse-traceless gauge \cite{Misner:1973}, a plane 
gravitational wave coming from direction $\hat n$ on the sky is given
by 
\begin{equation}
   h_{ij}(t, \vec{x}) = h_{+}(t,\vec x) \, e_{ij}^{+}(\hat n) 
            + h_{\times}(t,\vec x) \, e_{ij}^{\times}(\hat n)\,,
   \label{e:gwTT}
\end{equation}
where $h_{+,\times}(t,\vec x)=h_{+,\times}(t+\vec x\cdot\hat{n}/c)$,
and the polarisation tensors are
\begin{eqnarray}
   {e}^{+}_{ij}(\hat{n}) = {\ell}_i {\ell}_j -
                           {m}_i {m}_j , \\
   {e}^{\times}_{ij}(\hat{n}) = {\ell}_i {m}_j +
                                {\ell}_j {m}_i\,.
\end{eqnarray}
The unit vectors $\hat{\ell}$ and $\hat{m}$ are chosen so that
$\hat{\ell}, \hat{m}, \hat{n}$ form a right-handed orthonormal basis. 
The rotational degree of freedom associated with the choice of 
$\hat{\ell}$ and $\hat{m}$ in the plane perpendicular to $\hat{n}$ is 
often called the polarisation angle $\beta$. In what follows, we will
suppress $\beta$-dependence for simplicity.

Consider a Michelson interferometer with arms aligned along the 
unit vectors $\hat{a}$ and $\hat{b}$. In the long-wavelength
approximation, a signal produced by a gravitational wave in the
detector \cite{Thorne:1987, Schutz:1987} is given by
\begin{equation}
   V(t) = \frac{1}{2} (a^i a^j - b^i b^j) h_{ij}(t,\vec{0}) \,,
\end{equation}
where we assumed that the detector is located at $\vec{x}=\vec{0}$ and
its size is negligible. Equivalently, the signal can be written as 
\begin{equation}
   V(t) = F_{+}(\hat n) h_{+}(t) + F_{\times}(\hat n) h_{\times}(t)\,,
   \label{e:V(t)}
\end{equation}
where 
\begin{equation}
   F_{A}(\hat n) = \frac{1}{2} \, (a_i a_j - b_i b_j) 
   {e}_{A}^{ij}(\hat n)
   \label{e:FLW}
\end{equation}
are the interferometer responses to the two independent polarisations
$(A=+,\times)$ of the gravitational wave.  
In the frequency domain, (\ref{e:V(t)}) becomes
\begin{equation}
   \tilde V(f) = F_{+}(\hat{n}) \tilde{h}_{+}(f) + 
                 F_{\times}(\hat{n}) \tilde{h}_{\times}(f)\,.
   \label{e:VtildeLW}
\end{equation}
In what follows, tilde always denotes Fourier transform with respect
to $t$.

\begin{figure}
 \centering\includegraphics[width=\columnwidth]{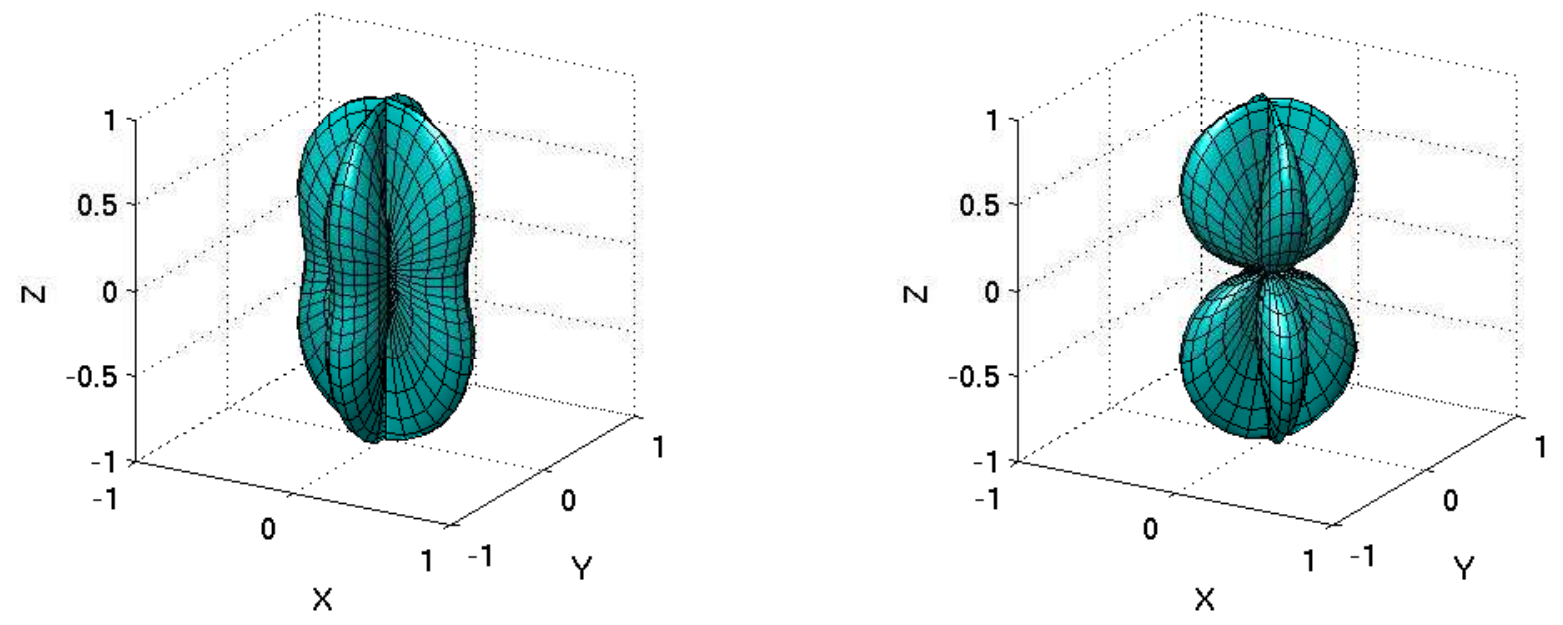}
   \caption{Antenna patterns in the long-wavelength approximation: 
   $|F_{+}(\hat{n})|$ (left) and $|F_{\times}(\hat{n})|$ (right). 
   They can also be obtained as the limiting case ($f=0$) of the exact
   response functions, described in section~\ref{s:Mich_cont}.}
   \label{fig:antennaDC}
\end{figure}

Three-dimensional representations of the absolute value of $F_{A}$ as
a function of $\hat{n}$ are often called \emph{antenna patterns}. 
Antenna patterns have traditionally been shown for a particular
choice of polarisation basis: $\hat{\ell}=\hat{\theta}$ and 
$\hat{m} = \hat{\phi}$, where $\hat{\theta}$ and $\hat{\phi}$ are
the unit vectors corresponding to the spherical coordinates 
$\phi\in[0,360^{\circ}]$ and $\theta\in[0,180^{\circ}]$.
Figure~\ref{fig:antennaDC} shows the antenna patterns in the coordinate
system with the $x$ and $y$ axes aligned with the interferometer
arms. In these coordinates,
$\hat{n}=(\sin\theta\cos\phi, \sin\theta\sin\phi, \cos\theta)$.

\section{Michelson interferometer response (exact formula)}
\label{s:michelson}

The detector response which takes into account the finite size of the
interferometer will be called here \emph{exact} in contrast to the
approximate response (\ref{e:FLW}). Here we give a brief derivation 
of the exact detector response following recent calculations in 
\cite{Rakhmanov:2006, Whelan:2007}.

\subsection{Photon propagation time}

The interval for photons propagating in spacetime with a 
gravitational wave (\ref{e:gwTT}) is 
\begin{equation}\label{eq:interval}
   \rmd s^2 = - c^2 \rmd t^2 + 
   \left[\delta_{ij} + h_{ij}(t,\vec x)\right] \rmd x^i \rmd x^j = 0 \,.
\end{equation}
Consider a photon launched in the direction $\hat{a}$ to be bounced 
back by a mirror some distance $L$ away. On the way forward, the 
unperturbed photon trajectory is $x^i = a^i \xi$, where $\xi\in[0,L]$.
Substituting this trajectory in (\ref{eq:interval}) and solving for
$t$, we obtain 
\begin{equation}
   c(t - t_0) = \int_{0}^{\xi} \left( 1 +
      h_{ij} \, a^i a^j \right)^{1/2} \rmd\xi' .
\end{equation}

Let $T$ be the nominal (unperturbed) photon transit time: 
$T \equiv L/c$. In the presence of a gravitational wave, the transit
time will slightly deviate from its nominal value giving rise to a 
small perturbation:
\begin{equation}
   \delta T(t) 
   = \frac{1}{2c} a^i a^j \int_0^L h_{ij} \left(t_0 + 
   \frac{\xi}{c} + \frac{\hat{n} \cdot \hat{a}}{c} \xi \right) 
   \rmd \xi ,
\end{equation}
where $t_0$ is the starting time for the photon propagation which can be
approximated by $t_0 = t - T$. Similarly, on the way back,
\begin{equation}
   \delta T'(t) 
   = \frac{1}{2c} a^i a^j \int_0^L h_{ij} \left(t_0 +
   \frac{L-\xi}{c} + \frac{\hat{n} \cdot \hat{a}}{c} \xi \right) 
   \rmd \xi ,
\end{equation}
where $t_0$ can also be approximated by $t_0 = t - T$. Then the
perturbation of the round-trip time is given by
\begin{equation}
   \delta T_{\mathrm{r.t.}}(t) = \delta T(t - T) + \delta T'(t) .
\end{equation}
In the Fourier domain, it can be written as 
\begin{equation}
  \frac{\delta \tilde{T}_{\mathrm{r.t.}}(f)}{T} = 
     a_i a_j \, D(\hat{a},f) \, 
     e^{ij}_{A}(\hat{n}) \, \tilde{h}_{A}(f) ,
  \label{e:single-arm}
\end{equation}
where we introduced the transfer function
\begin{equation}
  D(\hat{a},f) = 
  \frac{\rme^{-i2\pi fT}}{2}\,
  \left[
  \rme^{ i\pi f T_{+}} {\mathrm{sinc}} \left(\pi f T_{-} \right) +
  \rme^{-i\pi f T_{-}} {\mathrm{sinc}} \left(\pi f T_{+} \right) 
  \right] ,
  \label{e:D(a)}
\end{equation}
with short-hand notation: $T_{\pm} \equiv T(1 \pm \hat{a} \cdot \hat{n})$.
Further calculations require switching from photons to continuous
electro-magnetic waves.

\subsection{Phase lag of a continuous electro-magnetic wave}
\label{s:Mich_cont}

\begin{figure}
\centering\includegraphics[width=\columnwidth]{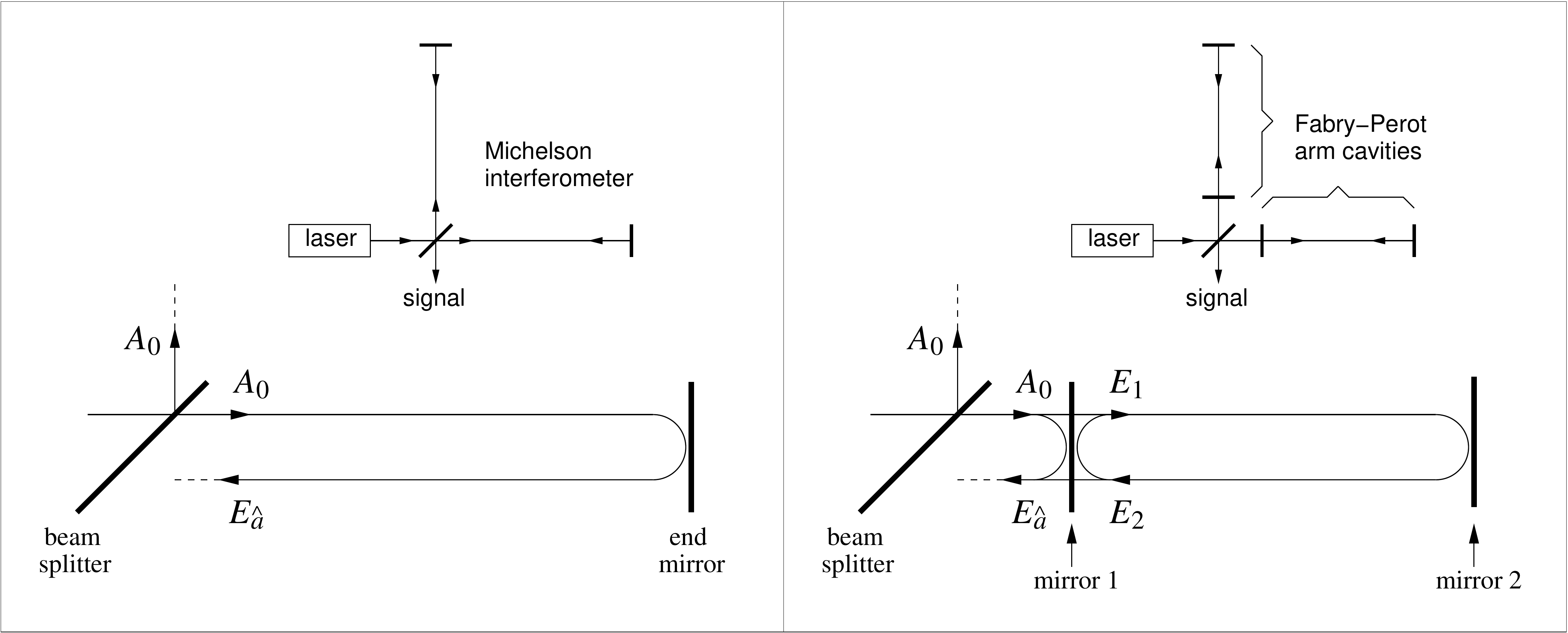}
   \caption{Simple Michelson interferometer (left) and Michelson 
   interferometer with Fabry-Perot arm cavities (right). Enlarged is 
   a schematic picture of the photon trajectory in the $\hat{a}$ arm. 
   (Forward and return paths are separated for clarity.) Here we 
   neglected the recycling mirror which increases the power incident 
   on the beam splitter but otherwise does not affect the detector 
   response.}
   \label{fig:ifo_configs}
\end{figure}

For an electro-magnetic wave propagating in the $\hat{a}$-direction, 
the electric field is given by 
$\mathcal{E}(t, \vec{x}) = A \exp[i(\omega t - k \vec{x}\cdot\hat{a})]$,
where $A$ is the amplitude, $\omega$ is the frequency, and $k$ is the 
wavenumber ($k=\omega/c$). It is convenient to suppress the
fast-oscillating factor $\rme^{i \omega t}$ by introducing the 
slowly-varying amplitude \cite{Rakhmanov:2002}: 
$E = \mathcal{E} \rme^{-i \omega t}$. 
Consider a simple Michelson interferometer with equal arm lengths,
$L$, as shown in figure~\ref{fig:ifo_configs} (left). The phase delay 
of the electro-magnetic field returning to the beam splitter after 
a round trip in the arm is  
$\omega (2T + \delta T_{\mathrm{r.t.}}) = 2kL + \psi$, where 
\begin{equation}
   \psi(t) = \omega \, \delta T_{\mathrm{r.t.}}(t) \,
   \label{e:phaseGW}
\end{equation}
is the phase delay due to the gravitational wave. Two such phases 
corresponding to arms $\hat{a}$ and $\hat{b}$ will be denoted here 
$\psi_{\hat{a}}$ and $\psi_{\hat{b}}$. Let the amplitude of the field 
immediately after the beam splitter be $A_0$. 
Then the amplitude of the field incident on the beam splitter from 
arm $\hat{a}$ is $E_{\hat{a}} = A_0 \, \exp(-2ikL -i\psi_{\hat{a}})$, 
and similarly for $E_{\hat{b}}$. Assuming that the interferometer  
operates at the dark fringe (destructive interference), we find that
the field at the output (signal) port is proportional to
\begin{equation}
   E_{\hat{a}}(t) - E_{\hat{b}}(t) \approx -i A_0 \, \rme^{-2ikL} 
   [ \psi_{\hat{a}}(t)  - \psi_{\hat{b}}(t) ] \,.
\end{equation}
With appropriate normalization\footnote{The normalization is such that
$\max_{\hat{n}}(G_A)=1$ at $f=0$ for both $A=+,\times$.} the signal 
is given by
\begin{equation}
   V(t) = \frac{1}{2 \omega T} 
        [ \psi_{\hat{a}}(t)  - \psi_{\hat{b}}(t) ] \, .
   \label{e:V_mich}
\end{equation}
In the Fourier domain, it can be written as 
\begin{equation}
   \tilde V(f) = 
   G_+(\hat{n},f) \tilde{h}_+(f) + 
   G_\times(\hat{n},f) \tilde{h}_\times(f)\, ,
   \label{e:Vtilde}
\end{equation}
where $G_A(\hat{n},f)$ are the exact detector responses to the two
independent polarisations of the gravitational wave:
\begin{equation}
   G_{A}(\hat{n},f) = 
      \frac{1}{2}
      \left[ 
      a_i a_j \, D(\hat{a}, f) - 
      b_i b_j \, D(\hat{b}, f) 
      \right] e^{ij}_{A} (\hat{n})\,.
      \label{e:michelson}
\end{equation}
Note that the long-wavelength formula (\ref{e:FLW}) is a special case
of the exact response:
\begin{equation}
   F_{A}(\hat{n}) = G_{A}(\hat{n}, 0) .
\end{equation}
As the frequency of the gravitational wave increases, the difference
between $F_A$ and $G_{A}$ becomes more and more pronounced 
\cite{Sigg:1997, Hunter:2005}. The most drastic change occurs at 
the inverse of the photon round-trip time: $f=1/(2T)$. In Fabry-Perot 
cavities this quantity is called the free spectral range or FSR (see
section~\ref{s:FP_resp}).

\begin{figure}
 \centering\includegraphics[width=\columnwidth]{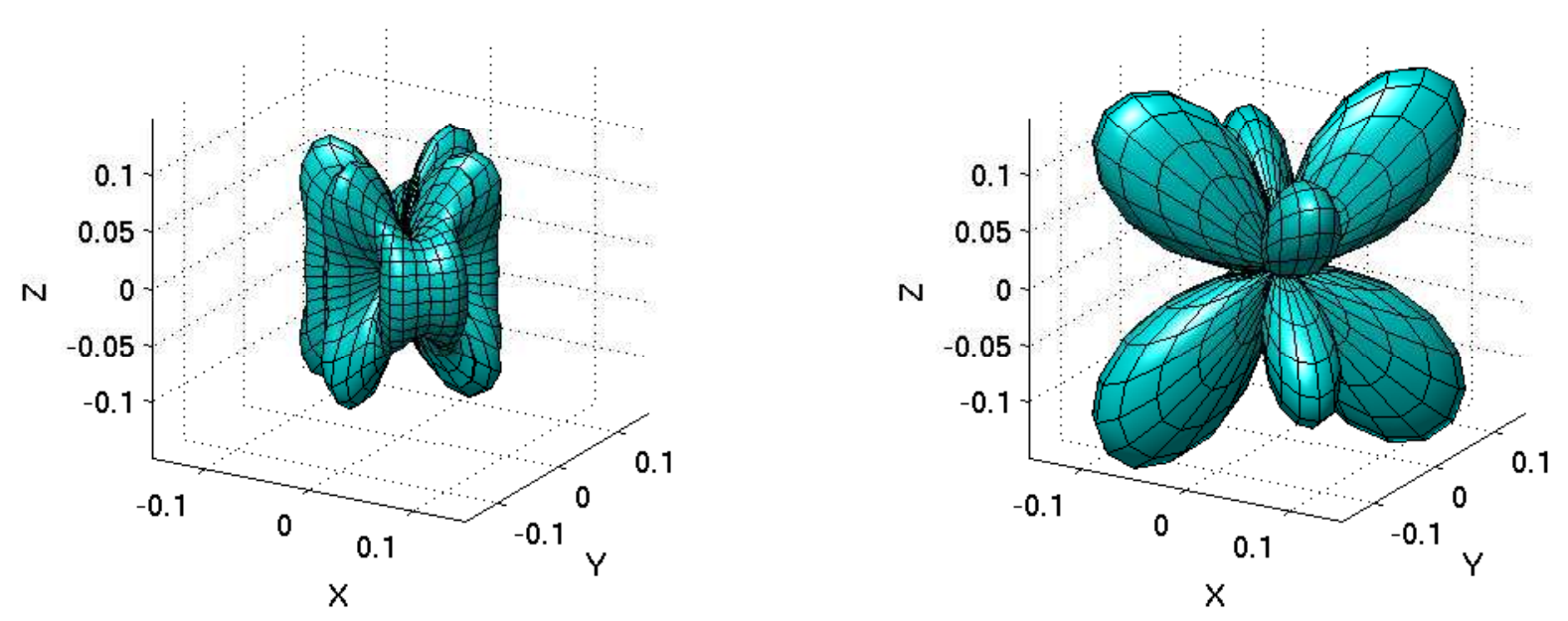}
   \caption{Antenna patterns at the FSR frequency (37.5~kHz): 
   $|G_{+}(\hat{n},f)|$ (left) and $|G_{\times}(\hat{n},f)|$ (right).}
   \label{fig:antennaFSR}
\end{figure}

Figure~\ref{fig:antennaFSR} shows the magnitude of the detector 
response functions $G_{A}(\hat{n},f)$ at the free spectral range of 
the 4-km LIGO interferometers ($f=37.5$~kHz). 
Note the presence of additional lobes in the antenna patterns, and 
a factor of 5-8 reduction in magnitude compared to the response at 
$f=0$, shown in figure~\ref{fig:antennaDC}.

\subsection{Approximate formulae for the Michelson response}

The long-wavelength approximation (\ref{e:FLW}) is obtained by
entirely neglecting the frequency dependence of the exact response 
\begin{equation}
   G_{A}(\hat{n}, f) \approx F_{A}(\hat{n}) .
   \label{e:zeroFreq}
\end{equation}
A moderate frequency dependence is obtained by adding the
first-order correction:
\begin{equation}
  G_{A}(\hat{n}, f) \approx F_{A}(\hat{n}) + 
      f\, \frac{\partial G_{A}}{\partial f} (\hat{n}, 0)\, .
      \label{e:linearFreq}
\end{equation}
This may not necessarily be a better approximation than 
(\ref{e:zeroFreq}) since the higher order terms neglected in
(\ref{e:linearFreq}) can be greater than the first-order term 
(proportional to first derivative). This happens because the 
first-order term vanishes at some locations in the sky whereas the 
sum of all higher order terms remains non-zero.

The most recent estimation of errors introduced by the long-wavelength
approximation is due to Baskaran and Grishchuk \cite{Baskaran:2004}. 
These authors argued in favour of the linear approximation of the
detector response (\ref{e:linearFreq}), and estimated the corrections 
to the long-wavelength formula using the first-order ($f$-proportional)
term. (In \cite{Baskaran:2004}, the two terms in the right-hand side 
of (\ref{e:linearFreq}) are called \emph{electric} and \emph{magnetic} 
components.) As we have shown, such a linear approximation is
problematic. Fortunately, one need not be concerned with the accuracy
of the linear approximation since the exact formula is readily available.

\section{Transfer function of Fabry-Perot arm cavities}

The response of LIGO and VIRGO detectors is enhanced by
incorporation of optical resonators (Fabry-Perot cavities) in
interferometer arms. Here we briefly derive the transfer function 
of a Fabry-Perot cavity and include it in the exact detector
response.

\subsection{Phase amplification due to multi-beam interference}
\label{s:FP_resp}

Consider a Fabry-Perot cavity in one of the arms of the
interferometer, as shown in figure~\ref{fig:ifo_configs} (right). 
Let the amplitude of the light incident on the cavity be $A_0$. 
Then the fields circulating in the cavity satisfy
\begin{eqnarray}
   E_1(t) & = & t_1 A_0 - r_1 E_2(t) , \label{e:E1} \\
   E_2(t) & = & - r_2 \, E_1(t - 2T) \, \rme^{-2ikL - i \psi(t)} ,
   \label{e:E2}
\end{eqnarray}
where $t_1$ is the transmissivity of the front mirror, and $r_{1,2}$
are the reflectivities of the front and back mirror, respectively. 
Here $\psi$ is the phase lag due to the gravitational wave, 
(\ref{e:phaseGW}). The condition for resonance implies that $L$ is
equal to an integer number of half-wavelengths of light and therefore
$\rme^{-2ikL}=1$.

The field returning to the beam splitter from the cavity, 
$E_{\hat{a}} = r_1 A_0 + t_1 E_2$,
consists of the promptly reflected field and the leakage field. 
The information about the gravitational wave 
is contained in the leakage field which is proportional to internal
field $E_2$. Let the amplitude and the phase of this field be $A$ and 
$\Psi$, i.e. $E_2 = A \rme^{-i\Psi}$. Solving equations (\ref{e:E1})
and (\ref{e:E2}) to first order in $\Psi$, we find that the amplitude 
is given by $A = - t_1 r_2 A_0/(1 - r_1 r_2)$ and that the phase
satisfies the equation
\begin{equation}
   \Psi(t) - r_1 r_2 \Psi(t - 2T) = \psi(t)\,,
   \label{e:PsiIterationEq}
\end{equation}
or equivalently,
\begin{equation}
   \Psi(t) = \sum_{k=0}^{\infty} (r_1 r_2)^k \psi(t - 2kT)\,.
   \label{e:PsiSummationEq}
\end{equation}
Taking the Fourier transform of either (\ref{e:PsiIterationEq}) or
(\ref{e:PsiSummationEq}), we obtain
\begin{equation}
   \tilde{\Psi}(f) = g_0 C(f) \tilde{\psi}(f) \, ,
   \qquad 
    C(f) = \frac{1 - r_1 r_2}{1 - r_1 r_2 \rme^{-i4\pi fT}}\, ,
   \label{e:H_FP}
\end{equation}
where $g_0=(1 - r_1 r_2)^{-1}$ is the cavity amplification factor
and $C(f)$ is the normalized transfer function. Note that $C(f)$ is 
a periodic function of frequency with the period known as
the free spectral range, $\mathrm{FSR}=1/(2T)$. The 4-km LIGO
interferometers have the cavity gain of 70.6 and the FSR of 37.5~kHz.

It is convenient to represent $C(f)$ in the following equivalent form:
\begin{equation}
   C(f) = \rme^{i 2\pi fT} \frac{\sinh(2\pi f_0 T)}
          {\sinh[2\pi f_0 T(1 + i f/f_0)]}\,,
          \label{e:H_FP_2}
\end{equation}
where $f_0$ is the lowest order pole, $f_0 = |\ln(r_1 r_2)|/(4 \pi T)$.
At low frequencies ($f \ll {\mathrm{FSR}}$), one can approximate the exact
response (\ref{e:H_FP_2}) with a zero-pole filter, 
\begin{equation}
   C_{\mathrm{zp}}(f) = \frac{1 + i f/f_1}{1 + i f/f_0} \, ,
   \label{e:Capprox}
\end{equation}
where $f_1 = {\mathrm{FSR}}/\pi$ is the frequency of the zero. In  
the 4-km LIGO interferometers, $f_0 = 85.1$~Hz and $f_1 = 11.9$~kHz.

\subsection{The response of a Michelson-Fabry-Perot interferometer}

Calculating the field at the output (signal) port, as we did 
in section~\ref{s:Mich_cont}, we find that the signal in a
Michelson interferometer with Fabry-Perot arm cavities is 
\begin{equation}
   \tilde V(f) = H_+(\hat{n},f) \tilde{h}_+(f) + 
   H_\times(\hat{n},f) \tilde{h}_\times(f) ,
   \label{e:Vtilde2}
\end{equation}
where $H_{A}$ can be found by combining the 
Michelson response (\ref{e:michelson}) with the transfer 
function of a Fabry-Perot cavity (\ref{e:H_FP}):
\begin{equation}
  H_{A}(\hat{n},f) = C(f) \, G_A(\hat{n}, f) \, .
  \label{e:exactRespMFP}
\end{equation}
Here we omitted the cavity gain ($g_0$) to have the detector response 
normalized to $1$ at $f=0$.

An ad-hoc low-frequency approximation for this formula is obtained 
by replacing the exact Michelson response, $G_A(\hat{n}, f)$, with its 
long-wavelength counterpart, $F_A(\hat{n})$, and by replacing the
exact Fabry-Perot response, $C(f)$, with the single-pole transfer
function,
\begin{equation}
   C_{\mathrm{pole}}(f) = \frac{1}{1 + if/f_0} \, .
   \label{e:Cpole}
\end{equation}
The result is the long-wavelength approximation for a
Michelson-Fabry-Perot interferometer:
\begin{equation}
   H_{A}^{\mathrm{approx}}(\hat{n},f) = 
   C_{\rm pole}(f)\, F_A(\hat{n}) \, ,
   \label{e:approxRespMFP}
\end{equation}
which is frequently used in theoretical calculations and data-analysis
algorithms.

\begin{figure}
\centering\includegraphics[width=\columnwidth]{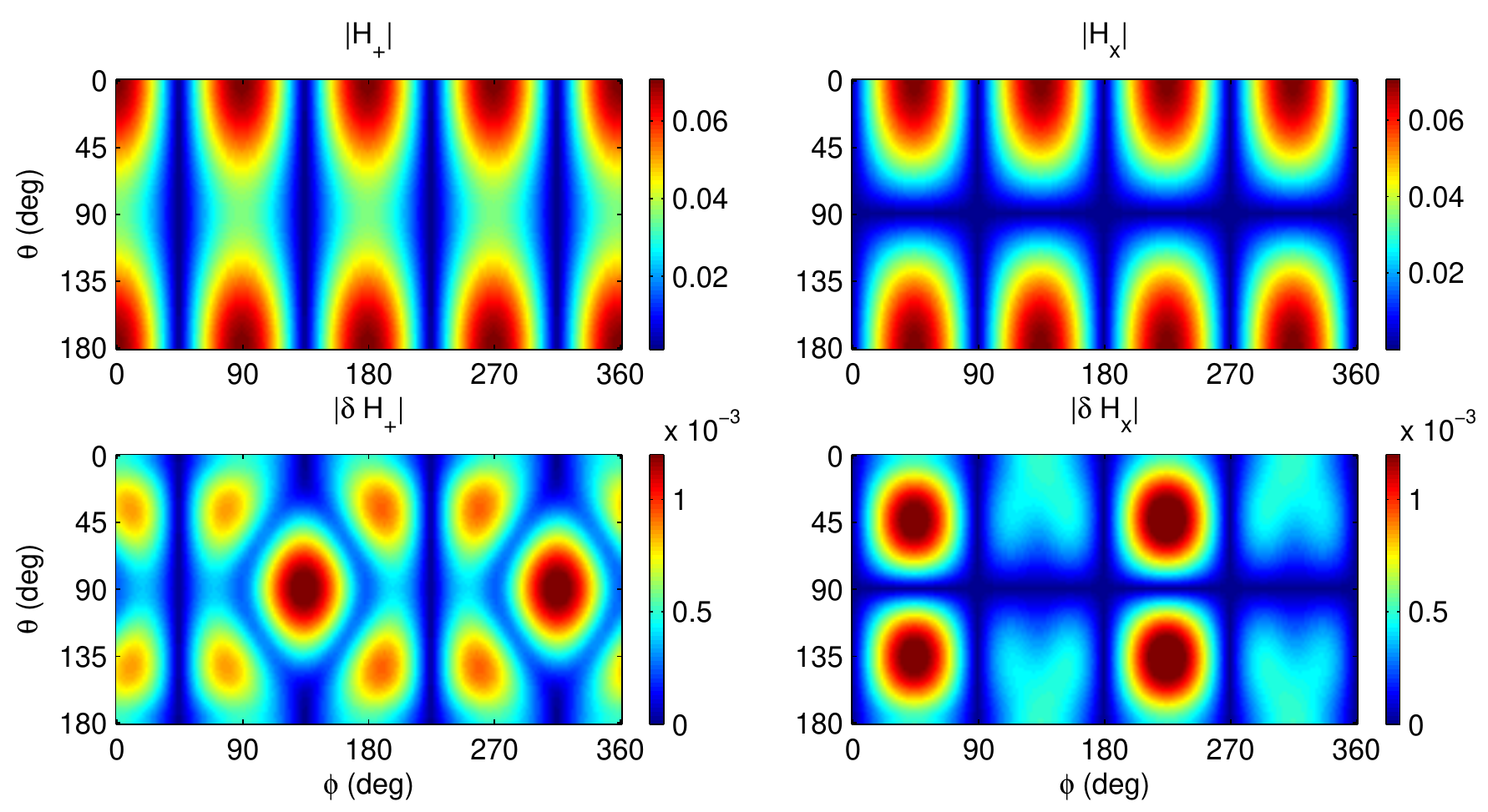}
   \caption{Top row: the normalized detector responses to $+$ and 
   $\times$ polarisations, bottom row: the magnitude of the error. 
   All these quantities are calculated at 1.2~kHz for comparison 
   with \cite{Baskaran:2004}.} 
   \label{fig:antennaCorr}
\end{figure}

The difference between the exact (\ref{e:exactRespMFP}) and 
approximate (\ref{e:approxRespMFP}) detector response, $\delta H_{A}$,   
is a source of systematic errors. The magnitude of $\delta H_A$ at a 
given frequency is a function of sky location, as shown in 
figure~\ref{fig:antennaCorr}. Note that the relative error, 
$\epsilon_A = |\delta H_A|/|H_A|$ diverges at those places in the 
sky where $H_A$ vanishes and $\delta H_A$ remains non-zero. Since 
the signal also vanishes at these locations, we can safely exclude 
all such places from the error estimation. Therefore, we consider 
only those locations in the sky where $|H_A|$ is greater than a 
certain value (threshold). For a conservative threshold of 25\% 
of the maximum value, we find that $\epsilon_A$ is at most 6-7\%. 
With a more realistic threshold of 50\%, $\epsilon_A$ is at most
2-3\%.

\subsection{Variations of the interferometer arm lengths}

In the following analysis, we will also need to consider the response
of the interferometer to changes of its arm lengths. Such a detector 
response is commonly used in calibration measurements. If $x_1$ and 
$x_2$ are the displacements of the front and back mirrors of one of 
the arm cavities, the change in the distance between the mirrors, as 
seen by the light propagating in the cavity, is
\begin{equation}
   \delta L(t) = x_2(t - T) - x_1(t)\, ,
   \label{e:lengthDef}
\end{equation}
where $T$ is the photon transit time ($T=L/c$). In this case, the 
signal\footnote{Consistency with a gravitational-wave signal requires 
that $\mu = 2/L$.} is
\begin{eqnarray}
   \tilde V(f) & = & \mu \, C(f) \delta \tilde{L}(f) \nonumber \\
   & = & \mu \,  C(f) \left[ \rme^{-i 2 \pi f T} \tilde{x}_2(f) - 
        \tilde{x}_1(f) \right] ,
   \label{e:calibExact}
\end{eqnarray}
where $C(f)$ is given by (\ref{e:H_FP}). (For derivation of 
(\ref{e:lengthDef}) and (\ref{e:calibExact}), see 
\cite{Rakhmanov:2002}.) At low frequencies, this response 
can be approximated by 
\begin{equation}
\tilde V(f) \approx \mu \left[
  C_{\mathrm{pole}}(f) \, \tilde{x}_2(f)-
  C_{\mathrm{zp}}(f)   \, \tilde{x}_1(f) \right] .
   \label{e:calibApprox}
\end{equation}
The relative error between the exact (\ref{e:calibExact}) and 
approximate formula (\ref{e:calibApprox}) is less than 1\% for the 
front mirror and 0.5\% for back mirror for frequencies up to
2~kHz.

\section{Effect on searches for periodic gravitational waves}
\label{s:periodic}

A gravitational-wave signal from a pulsar is quasi-periodic and 
therefore greatly benefits from synchronous detection (heterodyne 
method) \cite{Dupuis:2005}. This method removes the dominant
oscillatory part of the signal at frequency $f$, and corrects for 
any phase modulation (Doppler shift) due to the rotation of the 
Earth and its orbital motion relative to the pulsar, as well as 
any possible pulsar frequency evolution (e.g., spin-down).
If we ignore variations in the detector response due to these
small frequency shifts, the output of the heterodyne method is given by 
the following (complex) signal
\begin{equation}
   y(t) = \frac{1}{2} \left[ 
      H_{+}(\hat{n},f;t) \frac{1}{2}(1 + \cos^2\iota) - 
      i H_{\times}(\hat{n},f;t) \cos\iota \right] h_0 e^{i\phi_0},
   \label{e:pulsarEX}
\end{equation}
where $h_0$ is the amplitude, $\iota$ the inclination angle, and 
$\phi_0$ the initial phase of the heterodyne transformation. There 
is also an implicit dependence on the polarisation angle $\beta$ 
which enters the signal through the detector response functions. The
remaining time dependence in (\ref{e:pulsarEX}) comes from the
sidereal rotation of unit vectors $\hat{a}$ and $\hat{b}$ which are 
fixed to the Earth.

The use of the long-wavelength approximation affects both detection 
sensitivity and parameter estimation. For simplicity, assume that all 
parameters of the pulsar signal are known except for the amplitude
$h_0$. Then the relevant quantities are
\begin{equation}
   \epsilon = 1-
      \frac{ \langle y, z \rangle}
     {\sqrt{\langle y, y \rangle} \,
      \sqrt{\langle z, z \rangle}}
      \quad {\mathrm{and}} \quad
      \delta = 
      \frac{\langle y, z \rangle}{\langle z, z \rangle} - 1 \,,
   \label{e:delta}
\end{equation}
where $y(t)$ is given by (\ref{e:pulsarEX}) and $z(t)$ is calculated 
with the same expression except that $H_A$ are replaced with 
$H_A^{\mathrm{approx}}$ given in (\ref{e:approxRespMFP}). The inner 
product is defined as
\begin{equation}
   \langle y_u, y_v \rangle =  
   \int {\mathrm{Re}} \left[ y_u(t) y_v^*(t) \right] \rmd t \, ,
\end{equation}
where the integration is taken over one sidereal day. It can be shown 
that $\epsilon$ is the fractional change in signal-to-noise ratio, 
and $\delta$ is the fractional bias in the estimate of the amplitude 
of the gravitational wave, both caused by the use of the inaccurate 
detector response \cite{Woan:2008}.

\begin{figure}
\centering\includegraphics[width=\columnwidth]{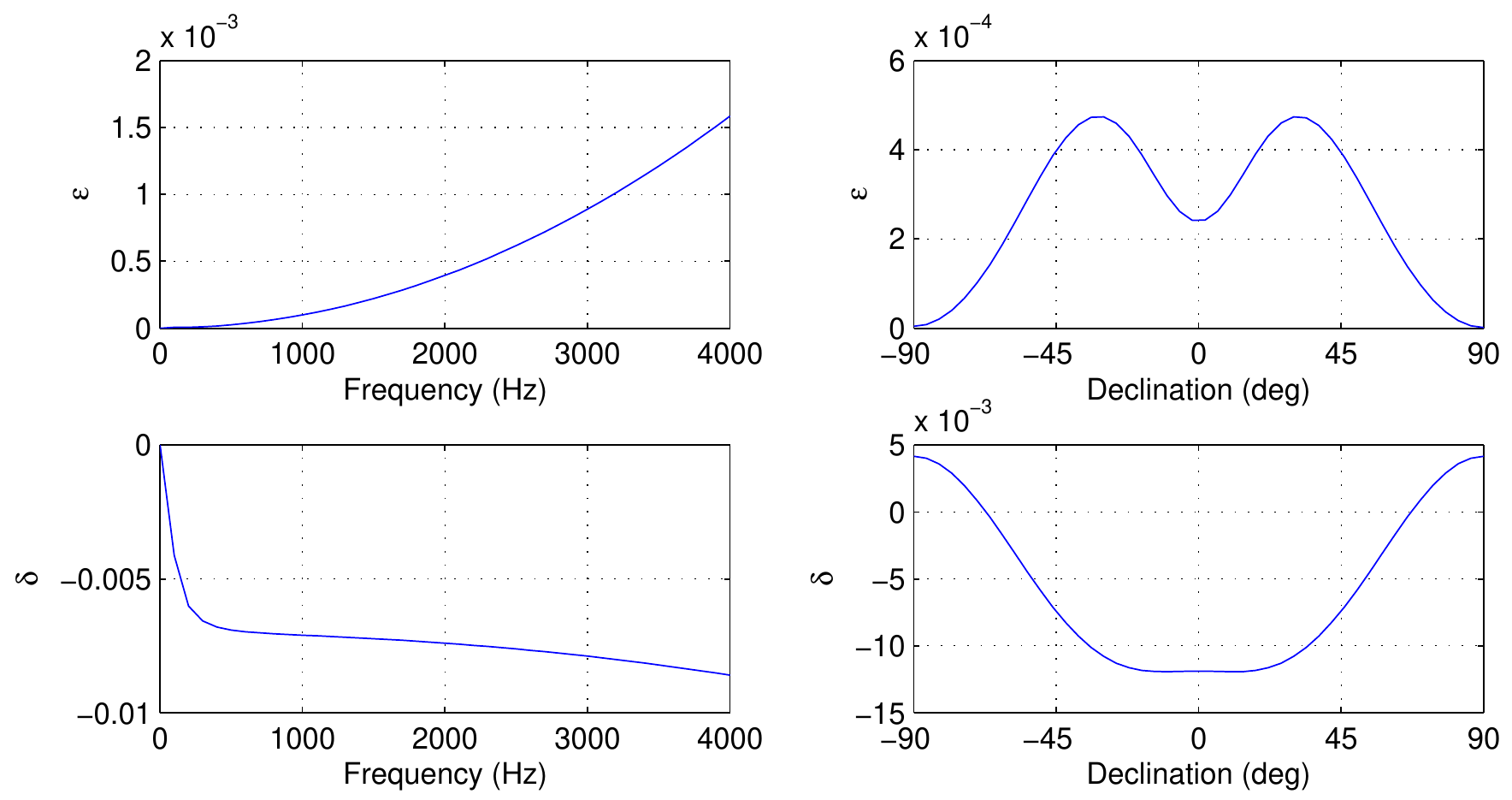}
   \caption{$\epsilon$ and $\delta$ as a 
   function of frequency for ${\mathrm{dec}}=45^\circ$ (left), 
   and as a function of declination for $f=2$~kHz (right). 
   (Other parameters are $\beta=0$, $\iota=0$, and $\phi_0=0$.)}
   \label{fig:match}
\end{figure}

Figure~\ref{fig:match} shows $\epsilon$ and $\delta$ as a function of 
the frequency of the gravitational wave for fixed source declination, 
and also as a function of the declination angle for fixed frequency, 
for the 4-km LIGO Hanford interferometer. For source location, we only 
need to specify the declination angle as the dependence on the right 
ascention is removed by the 24-hour sidereal-time integration. For 
simplicity, we assumed that $\beta=\iota=\phi_0=0$, which corresponds 
to a circularly-polarised gravitational wave. One can see from the 
figure that the change in the snr, $\epsilon$, is much less than 
1\% for all directions on the sky and all frequencies up to 4~kHz. 
Also, the bias, $\delta$, in the estimation of $h_0$ is 
less than 1.5\% for all directions on the sky at $f=2$~kHz.

\section{Effect on searches for stochastic gravitational waves}
\label{s:stochastic}

\begin{figure}
\centering\includegraphics[width=\columnwidth]{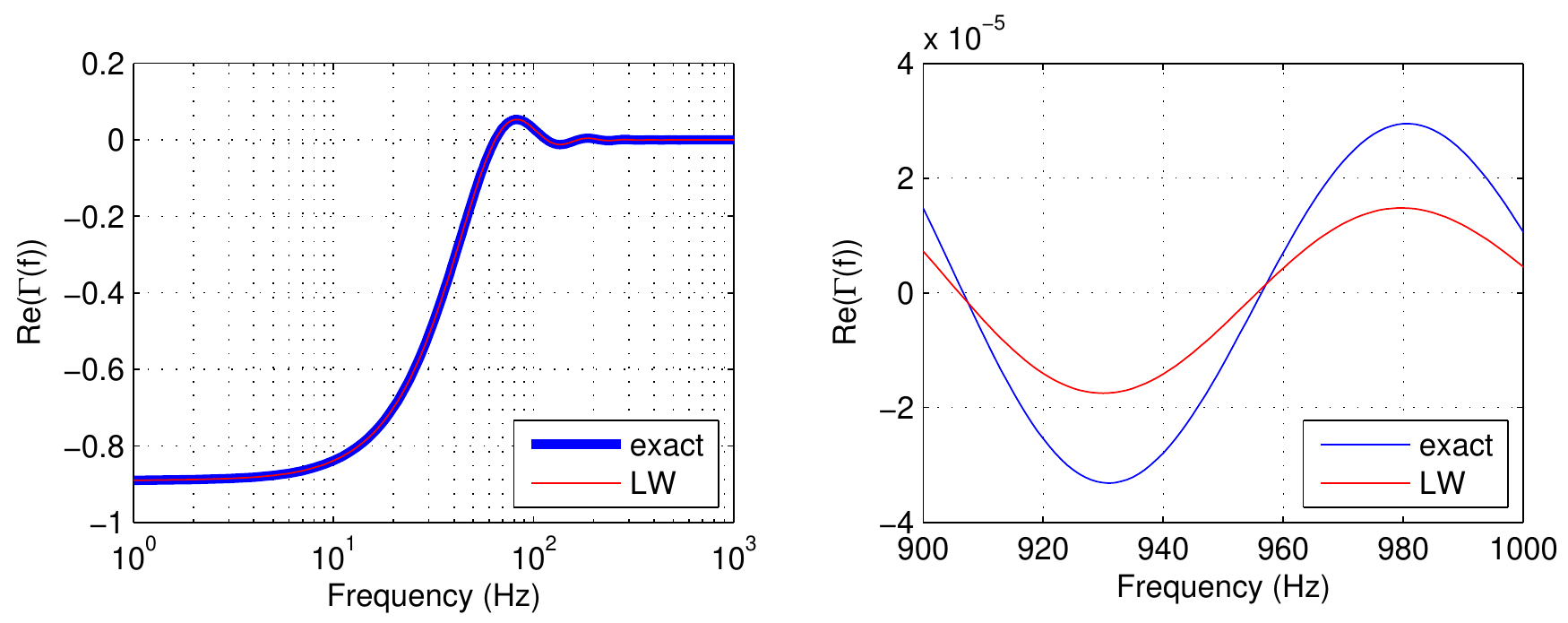}
\caption{Overlap reduction function for 
  LIGO H1 and L1 interferometers, calculated using the 
  long-wavelength and exact detector response functions.  
Left: ${\rm Re}(\Gamma(f))$ for frequency range from 1~Hz to 1~kHz. 
Right: a small section of the plot magnified to show the slight difference
  between the two curves.}
  \label{fig:GammaHL}
\end{figure}

Consider an isotropic, unpolarised stochastic gravitational-wave 
background and assume that it is described by a stationary Gaussian 
random process. Then the expectation value of the cross-correlation 
\cite{Christensen:1992, Flanagan:1993, Cornish:2001} of the outputs of
two detectors 
is proportional to the \emph{overlap reduction function}:
\begin{equation}
  \Gamma(f)=
  \frac{5}{8\pi} \int_{S^2} \rmd^2 \Omega_{\hat{n}}
  H_{1A}(\hat n,f)
  H^*_{2A}(\hat n,f)
  e^{i2\pi f\hat n\cdot(\vec{x}_1-\vec{x}_2)/{c}} \,,
  \label{e:overlapHH}
\end{equation}
where $\vec x_1$ and $\vec x_2$ are the locations of the two detectors
on Earth and $H_{1A}(\hat n,f)$ and $H_{2A}(\hat n,f)$ are their response 
functions. (The summation over polarisation indices $A$ is understood.) 
Figure~\ref{fig:GammaHL} shows a typical $\Gamma(f)$ calculated 
with both the exact (\ref{e:exactRespMFP}) and approximate
(\ref{e:approxRespMFP}) formulae. The two detectors are the 4-km 
LIGO Hanford (H1) and Livingston (L1) interferometers.

The error in the detector response from the long-wavelength 
approximation affects detection sensitivity and parameter estimation 
via the overlap reduction function. The two relevant quantities are
\begin{equation}
  \epsilon = 1-
    \frac{\langle \Gamma, \Gamma_{\ell.w.} \rangle}
  { \sqrt{\langle \Gamma, \Gamma \rangle} 
    \sqrt{\langle \Gamma_{\ell.w.}, \Gamma_{\ell.w.} \rangle} }
   \quad {\mathrm{and}} \quad
   \delta =   
    \frac{\langle \Gamma, \Gamma_{\ell.w.} \rangle}
 { \langle \Gamma_{\ell.w.}, \Gamma_{\ell.w.} \rangle} - 1 \, ,
\end{equation}
where $\Gamma(f)$ is given by (\ref{e:overlapHH}) and 
$\Gamma_{\ell.w.}(f)$ is calculated with the same expression except 
that $H_{A}$ are replaced with $H_{A}^{\mathrm{approx}}$ given in 
(\ref{e:approxRespMFP}). The inner product is defined as
\begin{equation}
\langle\Gamma_u, \Gamma_v\rangle
=
\int_{f_{\rm min}}^{f_{\rm max}}
\frac{{\rm Re}
   \left[
   \Gamma^*_u(f) \Gamma_v(f) 
   \right]}
   {f^6 Q_1(f) Q_2(f)}
   \rmd f\,,
\end{equation}
where $Q_{1,2}(f)$ are the power spectra of the outputs of the two
detectors. It can be shown that $\epsilon$ is the fractional change 
in signal-to-noise ratio, and $\delta$ is the fractional bias in 
the cross-correlation statistic, both caused by the use of the 
inaccurate detector response \cite{Woan:2008}. In this calculation, 
we also took into account the systematic error from the single-pole
approximation to the cavity response [see (\ref{e:calibApprox})] 
which occurs when the detector output is calibrated.

Table~\ref{tab:sigmacorr} shows $\epsilon$ and 
$\delta$ corresponding to several detector cross 
correlations. The upper part of the table gives the values 
for the LIGO-ALLEGRO search for a stochastic background
\cite{Abbott:2007}, which involved correlations of the Livingston 
interferometer with the ALLEGRO bar detector in a narrow band of 
frequencies near its peak sensitivity at 915~Hz. The search was
performed with three different orientations of the bar detector: 
parallel to the X-arm of the interferometer (AX), parallel to the 
Y-arm (AY), and parallel to the bisector of the two arms (AN),
also known as the {\em null} orientation. We find that the upper limits
in \cite{Abbott:2007} are not affected to the stated precision by 
the corrections in $\Gamma$.

\begin{table}
\caption{\label{tab:sigmacorr}
   The fractional snr change, $\epsilon$, and the fractional bias, 
   $\delta$, for an isotropic stochastic background. Here we assumed
   that the stochastic background has constant energy density in the 
   frequency band of interest $[f_{\rm min},f_{\rm max}]$, and used
   the nominal design sensitivities of the instruments. 
   (For details, see \cite{Woan:2008}.)
}
\begin{indented}
\lineup
\item[]\begin{tabular}{@{}ccccc}
\br
$f$ &  & \centre{3} {LIGO--ALLEGRO} \\
\mr
 $915$ Hz 
    &   $\qquad$  &   \quad {L1-AX} \quad   &   
                      \quad {L1-AY} \quad   &   
                      \quad {L1-AN} \quad  \\
\ns & \crule{4} \\
    & $\epsilon$  &  $\m3.6$e-7   &  $\m3.6$e-7   &  $\m3.6$e-7  \\
    & $\delta$    &   $-4.1$e-3   &   $-5.3$e-3   &   $-2.6$e-2  \\
\br
$[f_{\mathrm{min}}, f_{\mathrm{max}}]$  &   & \centre{3} {LIGO--VIRGO} \\
\mr
$50-150$ Hz   
    &             &  {\m H1-L1}   &  {\m H1-V1}   &  {\m L1-V1}  \\
\ns & \crule{4} \\
    & $\epsilon$  &  $\m4.0$e-6   &  $\m1.5$e-6   &  $\m1.5$e-6  \\
    & $\delta$    &   $-6.6$e-3   &   $-5.5$e-3   &   $-5.5$e-3  \\
\mr
$900-1000$ Hz  
    &             &  {\m H1-L1}   &  {\m H1-V1}   &  {\m L1-V1}  \\
\ns & \crule{4} \\
    & $\epsilon$  &  $\m1.2$e-3   &  $\m3.1$e-5   &  $\m2.2$e-5  \\
    & $\delta$    &  $\m2.7$e-2   &   $-1.5$e-2   &   $-1.5$e-2  \\
\br
\end{tabular}
\end{indented}
\end{table}

The rest of the table gives the values for $\epsilon$ and $\delta$ 
corresponding to various choices of cross-correlation between the 
LIGO and VIRGO (V1) interferometers. The first frequency band,
50-150~Hz, corresponds to the best sensitivity of the LIGO detectors.  
The second frequency band, 900-1000~Hz, is motivated by the VIRGO 
detector. At the present, this is where it contributes the most to 
the correlation-based searches. Note that 
only the H1-L1 low frequency analysis has been done so far, 
see e.g., \cite{Abbott:2007b}. The other interferometer
cross-correlations will be performed in the future.  
One can see from the table that all these errors 
are less than 1\%, except for the LIGO-VIRGO 
correlation searches around 1 kHz, which would 
have a 1-2\% fractional bias in parameter estimation.

\section{Summary}
\label{s:summary}

We re-evaluated high-frequency corrections to the detector response 
and analysed their effect on current searches for gravitational waves
with km-scale laser interferometers. Using the exact formula for the
detector response, we estimated systematic errors introduced by the
long-wavelength approximation in detection sensitivity and parameter
estimation of gravitational waves. Typical examples were taken from 
searches for periodic gravitational waves and from searches for an 
isotropic stochastic background. So far, in all cases considered, 
the errors are at most 1-2\% and somewhat smaller than the previously
reported 10\% error at 1.2~kHz \cite{Baskaran:2004}.

We have thus shown that the long-wavelength approximation for 
Michelson-Fabry-Perot interferometer (\ref{e:approxRespMFP}) was 
sufficiently accurate for searches performed to date, which were 
limited to frequencies below 2 kHz. However, extending the analysis 
to higher frequencies will likely require using the exact 
formula for the detector response (\ref{e:exactRespMFP}). 
For example, the exact formula is essential in searches for 
burst \cite{Parker:2007} and stochastic gravitational waves 
\cite{Forrest:2007} at the free-spectral-range frequency (37.5 kHz) 
of LIGO interferometers. Future searches for stochastic gravitational 
waves can go to even higher frequencies \cite{Nishizawa:2008}. 
In conclusion, we recommend the use of the exact formula whenever 
the accuracy of the detector response is important.

\ack

The authors would like to acknowledge discussions with D~Baskaran, 
N~Christensen, C~Cutler, S~Desai, L~Grishchuk, D~Khurana, G~Mendell, 
R~Prix, R~Savage, D~Sigg and G~Woan. This work was supported by the
Max-Planck-Society, by DFG grant SFB/TR~7, by the German Aerospace
Center (DLR), and by the US
National Science Foundation under grant PHY-0555842.
This article has been assigned LIGO Document Number P080036.

\section*{References}

\providecommand{\newblock}{}

\end{document}